\documentclass[11pt]{article}

\usepackage[utf8]{inputenc}
\usepackage[T1]{fontenc}
\usepackage{lmodern}
\usepackage{microtype}
\usepackage[margin=1.1in]{geometry}
\usepackage{graphicx}
\usepackage{booktabs}
\usepackage{amsmath}
\usepackage[round]{natbib}
\usepackage{xcolor}
\usepackage[colorlinks=true, linkcolor=blue!50!black, citecolor=blue!50!black, urlcolor=blue!50!black]{hyperref}

\title{The Calibration--Leverage Tradeoff in Exactly Solvable\\ Win-Probability Models}
\author{Devansh Mishra}
\date{July 2026}

\begin{document}

\maketitle

\begin{abstract}
We study ball-by-ball win probability (WP) for second-innings run chases in
Twenty20 cricket, built as an exactly solvable Markov model: we estimate a
single object, the per-ball outcome distribution over $\{0,\dots,6,
\text{wicket}\}$, and derive WP for every game state by backward induction
over the acyclic (balls, wickets, runs-required) chase graph. This
construction makes WP an exact martingale, which in turn makes \emph{leverage}
(how much a ball can swing WP) and \emph{win probability added} (WPA)
well-defined and exactly attributable; we use them to confirm that finishers
and death bowlers occupy the highest-leverage moments. We then show the
model's WP is systematically miscalibrated, and that this is not incidental.
Localizing the error, we rule out tail-thinning and marginal mis-estimation
(the model's per-ball outcome distribution matches the empirical one to a
total variation of $\le 0.02$ at every required run rate). The only remaining
cause is unmodelled dependence given the state, and we identify it: a
permutation-null decomposition shows short-range sequential run-scoring
persistence ($\sim$3--5 balls; innings-level heterogeneity contributes only
$\sim$18\%; wickets, if anything, anti-cluster). A block-bootstrap simulator
that injects the real dependence while holding the marginals fixed closes
26\% of the calibration gap (replicated over six seeds), saturating at block
lengths of $\sim$20 balls, consistent with the measured correlation range;
this is a constructive lower bound that confirms the diagnosis. The result is a structural
tradeoff relative to the (balls, wickets, runs) state description: exact
leverage requires the martingale, the martingale requires conditional ball
independence on that state, and that independence is what miscalibrates the
WP\@. Exactly attributable leverage and well-calibrated WP
cannot be obtained from the same object over this state.
\end{abstract}

\section{Introduction}\label{sec:intro}

Every ball of a Twenty20 run chase moves a probability. Broadcasters show it,
bettors trade it, and analysts increasingly use its movements to say which
players matter: how much a delivery \emph{could} swing the win probability is
the leverage of the moment, and how much a player's deliveries \emph{did}
swing it is his win probability added. Cricket has a distinguished line of
models for the probability itself, from the Duckworth--Lewis resource
valuations \citep{duckworth1998,stern2016} through dynamic-programming
treatments of limited-overs strategy \citep{clarke1988,preston2000} to WASP,
the ball-by-ball DP forecaster used in broadcast \citep{brooker2011}.
Baseball has a mature tradition for the attribution layer
\citep{studeman2004,tango2006}. This paper connects the two and reports
something that, to our knowledge, has not been stated: the property that
makes the attribution layer \emph{exact} and the property that makes the
probability layer \emph{correct} are, for this class of model, the same
assumption with opposite signs.

The model class is the natural one. For a second-innings chase the scoreboard
state (balls remaining, wickets in hand, runs required) describes the task
completely, and every legal delivery consumes a ball, so the state graph is
acyclic. Estimate one object, the distribution of the next ball's outcome
given the state, and win probability for \emph{every} state follows from a
single backward-induction sweep, exactly. Nothing about winning is fitted.
The construction has two consequences that flow in opposite directions, and
the paper is organized around the collision.

The first consequence is a gift. Constructed this way, win probability is a
martingale: from any state, the expected next-ball change is exactly zero.
Leverage becomes a well-defined conditional dispersion rather than a
heuristic, and win probability added becomes a fair game: a player
accumulates credit only by beating the state's expectation, weighted
automatically by the size of the moment. On eighteen seasons of Indian
Premier League chases this machinery confirms the sport's folk hypothesis
with clean accounting: leverage concentrates at the death ($2.7\times$ the
average ball in the final over), the batters who face the most of it are the
recognized finishers, and the clutch decomposition separates players who
perform in high-leverage moments from players who are merely present in them.

The second consequence is a debt, and quantifying it is the paper's main
work. The same model's probabilities are systematically
\emph{over-dispersed}: too confident at both extremes. Easy chases are
called safer than they are; hard-but-live chases are called more doomed (a
band the model rates at 18\% wins 29\% of the time), a shape that persists
out of sample. We localize the error and eliminate the candidate causes in
turn. It is not missing information: on the identical state, an unconstrained
learner beats the exact model by 0.094 nats, and the obvious latent-state
proxy, how long the striker has been at the crease, is a precise null. It
is not estimation error: the model's one-ball outcome distribution matches
the empirical one to total variation $\le 0.02$ everywhere. What remains,
necessarily, is the independence assumption itself, and we identify what it
discards. A permutation decomposition that provably absorbs innings-level
heterogeneity shows the dependence is \emph{sequential}: run-scoring persists
over a three-to-five-ball range, four-fifths of it beyond anything shared
innings conditions explain; wickets, against the folklore of collapse,
anti-cluster. A block-bootstrap simulator that injects this measured
dependence, with marginals held fixed and the experiment replicated across
seeds, closes 26\% of the calibration gap and saturates at the range the
decomposition measured.

The collision is then unavoidable, and we state it as the paper's thesis.
Exact leverage requires the martingale; the martingale requires conditional
independence of deliveries given the scoreboard state; and that independence
is what miscalibrates the probabilities. Within this state
description, exactly attributable leverage and calibrated win probability
cannot come from the same object, and the escape routes each surrender
something: enriching the state (the measured null, and a dependence with no
obvious pre-ball observable), fitting the probability directly (calibrated,
but it drifts along real paths and cannot host fair attribution), modelling
the dependence generatively (calibrated, but attribution becomes inference
over a hidden regime), or recalibrating post hoc (breaks the recursion
outright). We conclude with the practical resolution we adopt: two surfaces
with declared roles, attribution on the exact one with its error bound
published, probability reporting on a calibrated one.

Contributions, explicitly scoped:
\begin{enumerate}
  \item an exact chase win-probability engine with leverage and WPA layers,
    and the finisher/death-bowler hypothesis confirmed with exact accounting
    (Sections~\ref{sec:model}--\ref{sec:leverage}), including a
    state-conditional de-drifting correction that makes cross-role WPA totals
    comparable (Section~\ref{sec:wpa});
  \item a diagnosis of the exact model's miscalibration as over-dispersion,
    with marginal error eliminated and the residual pinned on conditional
    dependence (Sections~\ref{sec:validation}--\ref{sec:gap});
  \item a permutation decomposition separating sequential dependence from
    shared-latent heterogeneity (it finds short-range scoring persistence and
    wicket anti-clustering), and a replicated constructive experiment closing
    a quarter of the gap by injecting the measured dependence
    (Section~\ref{sec:dependence});
  \item the calibration--leverage tradeoff, stated relative to the scoreboard
    state, with each exit route priced (Section~\ref{sec:tradeoff}).
\end{enumerate}

We are explicit about what is \emph{not} claimed as new: the DP
win-probability construction is WASP's and ultimately Clarke's; leverage and
WPA are baseball's. The novelty is the diagnosis, the decomposition, the
constructive demonstration, and the tradeoff they jointly establish.

\section{Related work}\label{sec:related}

\paragraph{Win probability and state valuation in cricket.}
Dynamic programming enters limited-overs cricket with \citet{clarke1988}, who
solved optimal scoring rates over a (balls, wickets) state, with later
in-innings decision models in the same tradition
\citep{clarke1999,preston2000}. The Duckworth--Lewis method
\citep{duckworth1998} institutionalized the valuation of (overs, wickets)
resources, and Stern's update \citep{stern2016} recalibrated it to modern
scoring rates. This is the same era-shift problem our recency-weighted
estimation layer addresses (Section~\ref{sec:era}), solved there for target
resetting and here for probability calibration. WASP \citep{brooker2011} is the direct
ancestor of our substrate: a ball-by-ball DP model estimating expected runs
and win probability from historical transition frequencies, broadcast in New
Zealand from 2012. Our Section~\ref{sec:model} is this construction,
deliberately kept minimal; the paper's contribution begins where these models
stop, in asking what the construction's independence assumption does to its
own outputs.

\paragraph{Ball-by-ball simulators with player heterogeneity.}
The closest methodological neighbours are the Swartz-line simulators,
particularly \citet{davis2015}, whose T20 simulator draws each ball from
outcome distributions conditioned on batsman and bowler identities via
hierarchical empirical Bayes. That line and ours make opposite choices on the
same fork: they enrich the state with player latents and obtain richer
expectations by simulation, surrendering exactness; we hold the state to the
scoreboard and obtain exact, martingale-consistent attribution, surrendering
calibration, the cost this paper quantifies. Our ablation
(Section~\ref{sec:baselines}) and dependence decomposition
(Section~\ref{sec:decomposition}) can be read as measuring the cost of our
side of that fork, and Section~\ref{sec:escapes} as pricing theirs.

\paragraph{Leverage and clutch attribution.}
The leverage index originates with Tango \citep{tango2006,tango2007}, win
probability added with the sabermetric tradition around
\citet{studeman2004}; both are defined against empirical win-expectancy
tables. Transferring them to a \emph{derived}, exactly solvable WP changes
their epistemic status: leverage becomes the conditional MAD of a
constructed martingale rather than an estimate against a fitted surface.
That is why the calibration of the construction matters enough to occupy
the second half of this paper. In cricket, situational
pressure has been quantified by the pressure-index line of
\citet{bhattacharjee2016}; those indices are heuristic composites designed
for descriptive comparison, whereas our leverage is a model-derived quantity
with an exactness guarantee (and a measured bias).

\paragraph{Calibration and dependence.}
That unmodelled positive dependence over-disperses aggregate outcomes is
classical in statistics; our contribution is not the phenomenon but its
identification and constructive quantification inside a deployed model
class, and the observation that the independence being violated is
load-bearing for the model's attribution layer, not incidental. The
permutation device of Section~\ref{sec:decomposition} (within-group
shuffling as an exact null that preserves group composition while destroying
order) is standard machinery applied to a decomposition question we have not
seen posed in this setting: \emph{which kind} of dependence a
scoreboard-state model is missing.

To our knowledge, no prior work states the tension between exact
(martingale-based) attribution and probability calibration in sports win
models, measures both sides of it on one dataset, or demonstrates its
mechanism constructively. That is the gap this paper fills.

\section{An exactly solvable win-probability model}\label{sec:model}

\subsection{Data and the ball-level contract}\label{sec:data}

We use ball-by-ball data for the Indian Premier League, 2008--2026
(Cricsheet-derived), restricted to second innings: 1{,}162 chases, 130{,}029
legal deliveries. Chases are the deliberate choice of scope. With a fixed
target, the scoreboard state $s = (b, w, r)$ of legal balls remaining,
wickets in hand, and runs required is a complete description of the task,
and the terminal conditions are exact: $r \le 0$ is a win; $w = 0$ with runs
still required is a loss, as is $b = 0$ with $r > 1$; and finishing level
($b = 0$, $r = 1$) is a tie, valued $\tfrac12$.

Each delivery is reduced to a pre-ball state, one of eight outcomes
$o \in \{0, 1, 2, 3, 4, 5, 6, W\}$, and the realized match label
$y \in \{0, 1\}$. Three data-contract decisions are made once and disclosed
rather than modelled: extras are folded (a wide or no-ball adjusts runs
without consuming a legal ball), run-outs are folded into $W$, and matches
whose target changed mid-innings (rain-affected) are dropped. Monotonicity of
$b$, $w$, $r$ within innings and agreement of $y$ with the recorded winner
are enforced at ingestion.

The train/held-out split is temporal and fixed once for every experiment in
the paper: the three most recent seasons (2024--2026; 18{,}010 balls, 164
matches, realized win rate 0.498) are held out, and the remaining 998 innings
(112{,}019 balls) train every model.

\subsection{The recursion}\label{sec:recursion}

The single estimated object is the one-ball outcome distribution $p_o(s)$.
Everything else is arithmetic. Outcome $o = k$ runs sends $(b,w,r)$ to
$(b{-}1, w, r{-}k)$; a wicket sends it to $(b{-}1, w{-}1, r)$. Every legal
ball decrements $b$, so the transition graph is acyclic and finite, and win
probability is obtained in a single backward sweep:
\begin{equation}\label{eq:recursion}
  V(s) \;=\; \sum_o p_o(s)\, V\!\big(s'_o\big),
\end{equation}
with the terminal values above. There is no iteration and no fitting of $V$
itself. One structural consequence is immediate: $V$ satisfies the recursion
\emph{exactly} (machine-verified to $10^{-9}$ on random states), so under the
model the WP process is a martingale; from any state, the
probability-weighted WP of the next states equals the current WP\@.
Section~\ref{sec:leverage} builds on this property;
Sections~\ref{sec:gap}--\ref{sec:dependence} show its price.

\subsection{Estimating the outcome distribution}\label{sec:outcome}

The baseline estimator conditions $p_o$ on required run rate
($\mathrm{RRR} = 6r/b$, six bins), wickets in hand, and innings phase
(powerplay / middle / death), with two-level shrinkage: each cell shrinks
toward its (phase, wickets) parent, which shrinks toward the global
distribution, with Dirichlet-style pseudo-counts; unseen cells fall back down
the same chain. RRR and phase are estimation features only; the
dynamic-programming state remains $(b,w,r)$.

\subsection{The scoring-era problem, handled inside the estimation layer}\label{sec:era}

T20 scoring has drifted upward sharply: training seasons average 1.344 runs
per legal ball against 1.598 in the held-out 2024--2026 seasons, the
highest-scoring in league history. A model fit uniformly on 2008--2023
systematically under-predicts modern chases. We correct this \emph{inside the
estimation layer}, where it cannot damage the structure: each training ball
receives an exponential recency weight with half-life $h$ seasons, and every
level of the shrinkage hierarchy is fit from the same weights. The result is
still one outcome distribution and one backward sweep, so $V$ remains an
exact martingale. Post-hoc recalibration of $V$, by contrast, would break
the recursion (Section~\ref{sec:tradeoff}).

The half-life is tuned by rolling-origin cross-validation on the three most
recent \emph{training} seasons, each predicted one step ahead from strictly
earlier data: the same forecasting task the real held-out seasons pose, with
no contact with them. Because scoring trends monotonically, raw CV loss
decreases toward the aggressive end of the grid while the effective sample
size collapses; the selection rule, fixed in advance, takes the most
aggressive half-life on the improving curve with mean effective sample size
at least 15\% of the training data. This selects $h = 0.75$ seasons. On the
held-out seasons the era-adjusted model improves Brier from 0.1637 to 0.1489,
log loss from 0.5313 to 0.4741, and ECE from 0.1627 to 0.1310. This closes
63\% of the Brier gap to the strongest simple reference
(Section~\ref{sec:validation}) while keeping the martingale. The correction
has a hard ceiling: reweighting can pull the fitted scoring level only
toward the newest training season (weighted mean 1.424), not past it toward
the held-out 1.598; going further would require trend extrapolation, which we
decline.

\section{Leverage and clutch attribution on the exact surface}\label{sec:leverage}

\subsection{Definitions and validation}\label{sec:leverage-defs}

Because the model exposes the full next-ball distribution, not merely a win
probability, the \emph{swing} of a state is computable exactly:
\begin{equation}\label{eq:swing}
  \mathrm{swing}(s) \;=\; \sum_o p_o(s)\,\bigl|V(s'_o) - V(s)\bigr|,
\end{equation}
the conditional mean absolute deviation of the next-ball WP, following the
absolute-value convention of baseball's leverage index. The leverage index is
swing normalized by the average over the balls that actually occurred, so the
average real delivery has $\mathrm{LI} = 1$; the reference average is 0.0170
WP per ball. Two checks anchor the machinery. Analytically, in a synthetic
model where one ball decides the match as a coin flip, swing must approach
$2p(1-p) = 0.5$; the implementation returns 0.5000. Empirically, the
martingale property must hold along real sequences: the mean signed one-ball
WP change over all 130{,}029 balls is $+0.00048$, near zero in every WP
decile. The one dispersion diagnostic that \emph{fails} is residual lag-1
autocorrelation of runs within partnerships, $+0.037$ where the model implies
zero. We record it as the error bound on everything in this section; it is
the thread Section~\ref{sec:dependence} pulls.

\subsection{Who plays the high-leverage moments}\label{sec:who}

Leverage concentrates exactly where folk wisdom says: mean LI is 0.94 in the
powerplay, 0.79 in the middle overs, and 1.59 at the death, rising to 2.67 in
the final over; the highest-leverage individual balls are all tight
last-over chases. The batting-position profile is more interesting than the
folklore: mean LI \emph{rises} down the order from the openers (0.95) to a
peak at the finisher slots six and seven (1.31 and 1.26) and then collapses
for the genuine tail (0.24 at eleven). The tail bats at the death more than
anyone (death-ball share 0.88) yet faces the \emph{least} leverage, because
it bats in matches already decided. High death share does not imply high
leverage; a naive correlation over all eleven positions is negative
($-0.53$) purely from this artifact. Over the batting order proper (positions
1--7) the correlation is $+0.91$.

At the player level the hypothesis holds directly: the batters facing the
highest average leverage are the recognized finishers (H.~Pandya 1.40,
Dhoni 1.39, Pollard 1.34, all with career medians at positions 5--6), the
lowest are top-order anchors; and mean LI correlates with death-over share at
$+0.75$ for batters and $+0.56$ for bowlers. Who \emph{faces} the big moments
is a role, and the model measures it cleanly. Match-clustered intervals on
these leaderboards (Section~\ref{sec:limitations}) make the resolution
explicit: no adjacent pair in any top-ten table is statistically
distinguishable, while the top finisher and the lowest-leverage qualified
anchor differ in mean LI by $+0.69$ (95\% CI $[+0.33, +1.08]$). The rankings
separate roles, not neighbouring names.

\subsection{Win probability added, and making it fair}\label{sec:wpa}

WPA charges each ball's realized WP change to its participants: $\Delta V$ to
the batsman, $-\Delta V$ to the bowler, with the final ball stepping to the
realized $y$. Two accounting identities hold to machine precision and are
enforced by tests: WPA is zero-sum on every ball, and a side's WPA telescopes
over an innings to $y - V(\text{first ball})$. Because
$E[\Delta V \mid s] = 0$ under the model, positive WPA is earned only by
beating the state-conditioned expectation, automatically weighted by the
moment's leverage: the formal version of ``won the big moments.''

The martingale is exact under the model but the model is not calibrated
(Section~\ref{sec:gap}), and the aggregate consequence surfaces immediately:
summed over all innings, batting WPA is $+62.3$ wins rather than zero. This
is the model's calibration bias wearing the costume of skill: a baseline of
roughly $+0.00048$ WP per ball is credited to every batter and debited from
every bowler, and it invalidates raw cross-role comparisons. We remove it
\emph{state-conditionally}: each ball's WPA is de-drifted by the mean WPA of
all balls beginning in the same WP bin (width 0.02). The drift is far from
uniform, near zero in decided states and up to 0.0019 per ball in contested
ones, so a flat correction would mis-price players by where they operate. De-drifting preserves the per-ball zero-sum, drives the grand total
to zero by construction, and redistributes standings by state mix: the
largest individual effects are V.~Kohli falling from $+3.83$ to $+2.25$ (a
high-volume batter who faced more high-drift, contested balls than average)
and R.~Ashwin rising from $+4.21$ to $+5.55$.

The de-drifted clutch split is the section's payoff. Splitting each career at
$\mathrm{LI} = 2$, balls worth at least twice the average, separates
\emph{performing} from \emph{being present}: M.\,S.~Dhoni is essentially flat
on ordinary balls ($-0.001$ per ball) and strongly positive on his 346
high-leverage balls ($+0.017$ per ball), the statistical signature the word
``finisher'' gestures at; by contrast a genuinely excellent overall batter
can invert the split (S.\,A.~Yadav: $+0.003$ ordinary, $-0.010$
high-leverage). Match-clustered bootstrap intervals
(Section~\ref{sec:limitations}) put both named splits outside zero: Dhoni's
high-minus-low difference is $+0.018$ per ball (95\% CI $[+0.004, +0.032]$),
Yadav's $-0.013$ ($[-0.031, -0.001]$). One caution stands regardless:
per-ball differences below about $\pm 0.005$ sit inside the dependence error
bound of Section~\ref{sec:leverage-defs}, which these intervals condition on
rather than capture.

\section{How good is the exact WP? Validation and fitted baselines}\label{sec:validation}

\subsection{Protocol}\label{sec:protocol}

All models are scored on the identical held-out split by Brier score, log
loss, and expected calibration error, with reliability curves. Because all
$\sim$120 balls of an innings share one outcome label, per-ball resampling
understates uncertainty by roughly $\sqrt{120}$. \citet{brill2024} make this
point in general form: in a simulation study where the true win probability
is known, the clustered dependence of play-by-play data substantially
inflates the bias and variance of WP estimators, so honest intervals must be
far wider than per-event resampling suggests. Every model comparison
therefore reports a \emph{paired, match-clustered} bootstrap on the log-loss
difference (whole matches resampled, per-ball differences paired), and
``significant'' means the 95\% interval excludes zero.

\subsection{The exact model against simple references}\label{sec:refs}

\begin{table}[t]
\centering
\caption{Held-out comparison (18{,}010 balls, 164 matches). Markov and
logistic models are fit on the full training split; the XGBoost fits
surrender the 2023 season to early stopping.}
\label{tab:models}
\begin{tabular}{lccc}
\toprule
model & Brier & log loss & ECE \\
\midrule
constant base rate            & 0.2503 & 0.6937 & 0.0169 \\
base Markov (state-free $p_o$)& 0.2040 & 0.8011 & 0.2235 \\
RRR Markov                    & 0.1637 & 0.5313 & 0.1627 \\
RRR Markov + era ($h{=}0.75$) & 0.1489 & 0.4741 & 0.1310 \\
logistic on RRR alone         & 0.1404 & 0.4379 & 0.0773 \\
logistic, engineered features & 0.1350 & 0.4157 & 0.0700 \\
XGBoost on $(b,w,r)$          & 0.1414 & 0.4374 & 0.0771 \\
XGBoost, full features        & 0.1490 & 0.4607 & 0.0847 \\
\bottomrule
\end{tabular}
\end{table}

The exact model beats the trivial references soundly
(Table~\ref{tab:models}), and the era adjustment is a significant gain
($-0.057$ nats, CI $[-0.079, -0.035]$). But a \emph{one-feature} logistic on
required run rate alone beats the full backward-induction model. That
inversion is the paper's first symptom. The structured model is not short of
information; it conditions on strictly more than the logistic sees. It is
mis-mapping the information it has.

\subsection{Three results from the fitted baselines}\label{sec:baselines}

\paragraph{The gap is structure, not state.}
On the identical inputs $(b,w,r)$, an unconstrained gradient-boosted model
beats the exact Markov WP by 0.094 nats (CI $[0.026, 0.174]$), with ECE 0.077
against 0.163. Same information, different mapping. This figure is
conservative: the boosted models lose their most recent training season to
early stopping.

\paragraph{Hidden state is not the missing ingredient.}
Adding the striker's balls-faced count to the boosted model (the observable
proxy for a ``set'' batsman, and the enrichment the sport's folklore
recommends) moves held-out log loss by $+0.0004$ nats (CI $[-0.0032, +0.0035]$). A
null from the most flexible model we can fit is strong evidence: whatever the
exact model is missing, it is not carried by this variable.

\paragraph{Under distribution shift, rigidity is robustness.}
The best model on the held-out seasons is the plain logistic, beating every
boosted variant. The richest feature set is the \emph{worst} boosted model
because it can see the target total: it fits 2008--2023 scoring
conditions closely (best training loss, 0.410) and transfers worst (0.461),
and season-wise early stopping cannot catch this because the validation
season sits on the training side of the era shift. We flag this as a
practical warning for sports models generally: under a large covariate
shift, capacity converts into era overfitting.

Together, Table~\ref{tab:models} and the ablation frame the question the rest
of the paper answers. The exact model's deficit is not missing information
and not a missing latent variable; it is a property of the mapping from state
to probability. Section~\ref{sec:gap} locates that property.

\section{Localizing the calibration gap}\label{sec:gap}

Section~\ref{sec:validation} established that the Markov win probability is
miscalibrated in a way that additional state cannot repair: on the identical
state $(b,w,r)$, an unconstrained gradient-boosted model beats the
backward-induction WP by 0.094 nats of held-out log loss, while adding a
set-batsman proxy to that same unconstrained model moves log loss by
$+0.0004$ nats, a decisive null. The information in the state is not the
problem. Something about the \emph{mapping} from state to probability is.
This section localizes the error; Section~\ref{sec:dependence} identifies its
cause and reproduces it constructively.

\subsection{Design}\label{sec:gap-design}

Two decisions matter for what follows. First, all diagnostics in this section
are computed on the training split, scored by the model fit to that same
split. Section~\ref{sec:validation} showed that the held-out seasons sit on
the far side of a large scoring-era shift; evaluating train-on-train removes
that shift from the frame, so whatever miscalibration remains is structural
rather than temporal. (Section~\ref{sec:elimination} returns to the held-out
data to confirm the structure is not an in-sample artifact.) Second, the WP
surface under examination is the era-adjusted RRR Markov model of
Section~\ref{sec:model}: the surface on which the leverage and WPA results
of Section~\ref{sec:leverage} are actually computed, so any defect found here
propagates to those results directly.

We slice the state space by required run rate ($\mathrm{RRR} = 6r/b$), the
natural difficulty axis of a chase, using bins finer than the six the outcome
model conditions on, and compare the model's mean WP against the empirical
win rate within each slice:
\begin{equation*}
  G(\text{slice}) \;=\; \overline{V(s)}_{\,\text{slice}} \;-\;
  \overline{y}_{\,\text{slice}}.
\end{equation*}
A positive $G$ means the model overrates the chasing side in that band; a
negative $G$ means it underrates them.

\subsection{The gap flips sign: over-dispersion, not tail-thinning}\label{sec:signflip}

Entering this analysis, our working hypothesis was \emph{tail-thinning}: a
model that treats balls as independent draws should underweight the scoring
explosions that rescue difficult chases, and therefore rate hard chases as
harder than they are: a gap that is negative and grows monotonically with
RRR\@. The data reject it (Table~\ref{tab:gap}).

\begin{table}[t]
\centering
\caption{Calibration gap by required run rate, training split (112{,}019
balls). The gap changes sign: positive on easy chases, deepest in the live
middle, fading toward zero for near-hopeless chases.}
\label{tab:gap}
\begin{tabular}{lrccc}
\toprule
RRR slice & balls & mean model WP & empirical win rate & gap $G$ \\
\midrule
0--6   & 15{,}210 & 0.977 & 0.963 & $\mathbf{+0.014}$ \\
6--8   & 24{,}190 & 0.785 & 0.753 & $\mathbf{+0.032}$ \\
8--10  & 33{,}801 & 0.447 & 0.518 & $-0.071$ \\
10--12 & 18{,}194 & 0.183 & 0.294 & $\mathbf{-0.112}$ \\
12--14 & 8{,}034  & 0.080 & 0.165 & $-0.085$ \\
14--16 & 4{,}021  & 0.047 & 0.100 & $-0.054$ \\
16--18 & 2{,}263  & 0.028 & 0.062 & $-0.033$ \\
18--22 & 2{,}133  & 0.011 & 0.022 & $-0.011$ \\
22--40 & 2{,}652  & 0.006 & 0.014 & $-0.008$ \\
\bottomrule
\end{tabular}
\end{table}

The gap is not monotone in RRR (rank correlation with slice order: $-0.08$).
It is positive on the easiest chases (the model calls them \emph{safer} than
they are) and strongly negative in the contested middle, worst at RRR
10--12, where chases the model rates at 18\% are won 29\% of the time. This
is the signature of \emph{over-dispersion}: the model's probabilities are
pushed too far toward both 0 and 1. Outcomes are systematically less
determined than the model believes, in both directions.

This refines, and partly corrects, the summary a coarser analysis would give.
Averaged over all live balls, the model does under-predict (weighted mean WP
0.482 against an empirical 0.522), which reads as simple under-confidence.
The sign-flip shows why that reading is wrong: live balls merely cluster on
the hard side of the crossover. The defect is symmetric; the aggregate bias
is a composition effect.

Splitting Table~\ref{tab:gap} by wickets in hand adds one detail that matters
later. The positive band, the over-confidence on easy chases, lives almost
entirely in the 7--10-wickets-in-hand group ($+0.020$ and $+0.033$ in the two
easiest slices). At this point in the analysis, a natural story suggests
itself: the model, drawing each ball independently, cannot represent batting
collapses, so it misses the occasional easy chase that falls apart from six
down. Section~\ref{sec:dependence} tests that story directly, and it turns
out to be wrong in an instructive way.

\subsection{The marginals are not the problem}\label{sec:marginals}

One family of explanations must be eliminated before dependence can be
blamed: perhaps the one-ball distributions $p_o(s)$ are simply misestimated
where it matters. The outcome model bins RRR coarsely, with a single $15+$
catch-all shrunk toward calmer parent cells, so extreme chases are exactly
where estimation error should concentrate. If the model understates
$p_4 + p_6$ at high RRR, it mechanically understates the win probability
of hard chases.

We therefore compare the model's own per-ball probabilities against empirical
frequencies along the same fine RRR axis. For the scoring tail, the shortfall
$\overline{p_4 + p_6}^{\,\mathrm{emp}} -
\overline{p_4 + p_6}^{\,\mathrm{model}}$ never exceeds 0.006 in absolute
value in any slice; for the wicket probability the corresponding figure is
0.008. Both are at the noise level, and neither shows a trend in RRR\@.
(Section~\ref{sec:constructive} completes this check for the full
eight-outcome distribution; the conclusion is unchanged.) The model knows,
essentially exactly, how often each outcome occurs from every kind of
situation. Finer bins or lighter shrinkage would buy nothing.

\subsection{The elimination, and its out-of-sample check}\label{sec:elimination}

The two facts above force a conclusion by construction. If the one-ball
marginals are correct \emph{and} consecutive balls are conditionally
independent given the state, backward induction reproduces
$E[y \mid b,w,r]$ exactly; that is what the recursion computes. The
marginals are correct (Section~\ref{sec:marginals}), yet $V(s)$ misses
$E[y \mid s]$ with a systematic, sign-flipping shape
(Section~\ref{sec:signflip}). The only assumption left standing is
independence. The residual is \emph{conditional dependence between balls,
given the state}.

The same table computed on the held-out seasons confirms the shape is not an
artifact of in-sample fitting. There the era shift is back in frame, so the
level drops as expected (the deep middle trough reaches $-0.267$ at RRR
8--10), but the structure survives: still positive on the easiest chases
($+0.026$), still sign-flipping, still fading in the hopeless tail. Whatever
produces the over-dispersion is a property of the game, not of the fit.

What this section cannot do is say what the dependence \emph{is}. That
distinction, between two mechanisms observationally identical to every
diagnostic used so far, decides both the interpretation and the available
remedies. We turn to it next.

\section{The dependence: what it is, and that it closes the gap}\label{sec:dependence}

\subsection{Two mechanisms, one signature}\label{sec:two-mechanisms}

``Conditional dependence given the state'' admits two physically different
readings.

\emph{Serial correlation.} Outcomes influence nearby outcomes: a batter who
has just scored keeps scoring; boundaries and dot-balls arrive in runs. The
process has momentum that the state $(b,w,r)$ does not carry.

\emph{Shared latent heterogeneity.} Each innings (or partnership) has a
hidden level that all its balls share: pitch, ground dimensions, quality of
the bowling attack, who is batting. Conditional on the innings, balls
could be perfectly independent; pooling across innings still produces
dependence relative to a model that sees only $(b,w,r)$.

Every diagnostic so far is blind to the difference. Both mechanisms leave the
one-ball marginals intact; both over-disperse innings trajectories; both
produce positive residual autocorrelation; and both are carried by the
consecutive-ball blocks used in Section~\ref{sec:constructive}, since twenty
consecutive balls share their innings' latent as well as their local
ordering. The distinction nonetheless matters twice over. It decides what the
leverage and WPA results of Section~\ref{sec:leverage} mean (a set-batsman
momentum effect is chargeable to players in a way that pitch conditions are
not), and it decides the remedy (a latent innings effect suggests state
augmentation; short-range momentum does not correspond to any pre-ball
observable). It also matters for our own earlier claims: the residual lag-1
autocorrelation of $+0.037$ reported in Section~\ref{sec:leverage-defs} as a
``hidden-state error bound'' was computed after removing state means but not
innings effects, so it, too, conflates the two mechanisms.

\subsection{A permutation decomposition}\label{sec:decomposition}

The two mechanisms can be separated with an exact resampling argument.
Working with residuals
$x_t = \mathrm{runs}_t - \overline{\mathrm{runs}}_{(\mathrm{phase},\, w)}$
(state means removed, as in Section~\ref{sec:leverage}), consider shuffling
the residuals \emph{within each innings}: values permuted, positions fixed.
The shuffle destroys all ordering, and with it any serial structure. But it
preserves each innings' composition: any two balls of the same innings still
share that innings' latent level, so the shared-latent contribution to any
pair statistic survives the shuffle untouched. Repeating the shuffle builds a
null distribution with two useful properties:
\begin{enumerate}
  \item the null is \emph{centred at the heterogeneity component}, not at
    zero: the permuted autocorrelation is what innings-level sharing
    alone would produce; and
  \item an observed statistic \emph{above the null band} demonstrates
    dependence beyond innings-level heterogeneity, with the mechanical bias
    of any demeaning step absorbed into the null automatically.
\end{enumerate}

We validated the discriminator on synthetic data before applying it:
i.i.d.\ sequences are not flagged (null centred at zero, observed inside the
band); a pure innings-level random effect is fully absorbed (intraclass
correlation recovered, lag profile flat at that level, observed inside the
band at every lag); an AR(1) process is flagged with the correct decaying
profile.

\begin{figure}[t]
\centering
\includegraphics[width=0.85\linewidth]{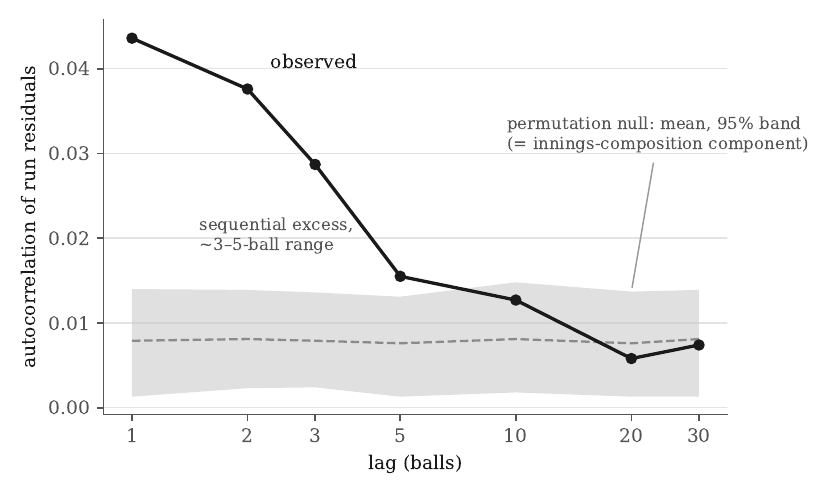}
\caption{Within-innings autocorrelation of state-demeaned run residuals,
against the within-innings permutation null (112{,}019 training balls, 200
permutations). The observed profile exits the flat null band at lags 1--5 and
decays into it by lag 10--20; the null band's centre is the
innings-composition (heterogeneity) component.}
\label{fig:lagprofile}
\end{figure}

Applied to the 112{,}019 training balls (200 permutations), the decomposition
gives an unambiguous answer (Figure~\ref{fig:lagprofile}). Three
observations, in increasing order of consequence.

\paragraph{The dependence is mostly not heterogeneity.}
The null centre at lag 1, the innings-composition component, is $+0.008$
against an observed $+0.044$: shared innings conditions account for about
18\% of the signal. The null centre agrees with the innings-level intraclass
correlation computed independently (0.0079 vs.\ 0.0081), an internal
consistency check on the whole construction.

\paragraph{The remainder is short-range and sequential.}
The excess over the null decays from $+0.036$ at lag 1 to $+0.021$ at lag 3
to $+0.008$ at lag 5, and is indistinguishable from the null by lag
10--20: a correlation range of roughly three to five balls. It survives the
demeaning ladder: removing innings means leaves lag-1 at $+0.027$ against a
null upper bound of $-0.003$; removing partnership means and restricting to
within-partnership pairs leaves the observed value above its (strongly
negative, bias-absorbing) null band as well. Dependence that survives
partnership demeaning cannot be an artifact of \emph{who} is batting or
\emph{how set} they are in the aggregate; it is ball-adjacent scoring
momentum.

\paragraph{Wickets do not cluster; they anti-cluster.}
Applying the identical machinery to wicket-indicator residuals reverses the
sign: at lags 1--3 the observed autocorrelation sits \emph{below} the
permutation band (lag 1: $-0.009$ against a null of $[-0.006, +0.005]$).
Immediately after a wicket, beyond what the state adjustment already
accounts for, another wicket in the next few balls is \emph{less} likely
than independence predicts. Batting sides consolidate. This kills the
collapse story suggested at the end of Section~\ref{sec:signflip}: the
over-confidence on easy chases is not the model missing wicket cascades. It
does not need to be. Positive scoring persistence alone produces the
two-sided gap, because persistence is symmetric: bursts rescue hard chases
(the model was too pessimistic there), and the mirror-image droughts sink
easy ones (too optimistic there).

\subsection{Constructive confirmation}\label{sec:constructive}

The elimination argument of Section~\ref{sec:elimination} and the
decomposition of Section~\ref{sec:decomposition} identify the cause but do
not yet demonstrate it. The demonstration should run forwards: inject the
measured dependence into an otherwise identical model and watch the
calibration gap close. We build a forward innings simulator with a dependence
dial.

\paragraph{Completing the marginal check.}
The simulator resamples real balls, so all comparisons require that the
model's marginals match the empirical ones in full, not only in the scoring
and wicket tails checked in Section~\ref{sec:marginals}. Total variation
distance between the model's eight-outcome distribution and the empirical one
is at most 0.018 across every RRR slice, and at most 0.0006 in the four most
heavily populated slices (RRR 0--12, 91{,}000 of the 112{,}019 balls), which
carry both signs of the flip and the worst single slice
(Table~\ref{tab:tv}). The elimination is complete at the level of the entire
one-ball distribution.

\begin{table}[t]
\centering
\caption{Full one-ball marginal check: total variation distance between the
model's eight-outcome distribution and the empirical one, by RRR slice
(training split).}
\label{tab:tv}
\begin{tabular}{lrcc}
\toprule
RRR slice & balls & TV distance & max.\ single-outcome deviation \\
\midrule
0--6   & 14{,}069 & 0.0006 & 0.0005 \\
6--8   & 24{,}793 & 0.0003 & 0.0002 \\
8--10  & 33{,}823 & 0.0002 & 0.0001 \\
10--12 & 17{,}747 & 0.0004 & 0.0003 \\
12--14 & 8{,}784  & 0.0038 & 0.0022 \\
14--16 & 4{,}087  & 0.0116 & 0.0090 \\
16--18 & 2{,}101  & 0.0141 & 0.0128 \\
18--22 & 2{,}370  & 0.0181 & 0.0128 \\
22--40 & 4{,}245  & 0.0117 & 0.0095 \\
\bottomrule
\end{tabular}
\end{table}

\paragraph{Modes.}
From each evaluated start state we run Monte-Carlo continuations to
termination under three regimes. \emph{model-iid} draws every ball
independently from the fitted $p_o(s)$; since this reproduces exactly the
process backward induction integrates, its mean terminal value must equal the
DP win probability, and it does. This validates the simulator's mechanics.
\emph{block-$K$} resamples runs of $K$ consecutive real balls, matched to the
current situation cell at each block start: the per-ball marginals stay flat
by construction, but real local dependence up to range $K$ rides along.
$K=1$ is an independent draw from the empirical cell marginal and must
agree with model-iid; it does. Raising $K$ raises only one thing: the amount of
real sequential structure the simulation honours.

\paragraph{Protocol.}
400 continuations per state, 220 evaluated states per RRR slice, and the
entire experiment replicated over six seeds, each seed redrawing both the
evaluated states and the Monte-Carlo paths. The calibration criterion is the
mean absolute slice gap, $\overline{|G|}$. Because the resampler uses flat
empirical frequencies, this experiment uses the plain (non-era-weighted) fit
throughout, so that all modes share the same marginals; its independence
references reproduce the same gap structure as Section~\ref{sec:signflip}.
As a control, each mode's \emph{realized} outcome distribution (what its
simulations actually consumed) is compared against the $K{=}1$ mode's:
block modes stay within total variation 0.0033 of independence across all
seeds, so any gap movement is attributable to dependence alone.

\begin{figure}[t]
\centering
\includegraphics[width=0.85\linewidth]{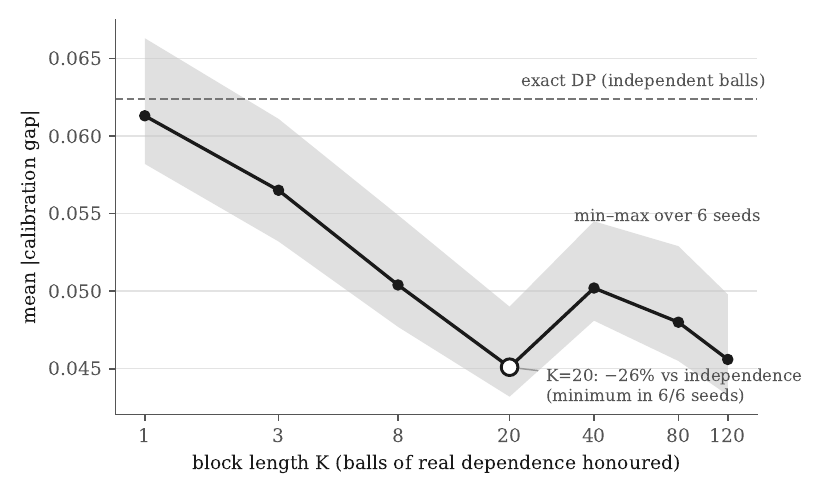}
\caption{Mean absolute calibration gap by dependence range $K$, replicated
over six seeds (band: min--max across seeds). Injecting real dependence with
marginals held fixed closes the gap monotonically out to $K=20$ (a 26\%
reduction relative to independence, with the minimum's location stable in six
seeds of six) and saturates thereafter.}
\label{fig:kcurve}
\end{figure}

Injecting real dependence with marginals held fixed closes the gap
monotonically out to $K=20$: from 0.0613 to 0.0451, a 26\% reduction
relative to independence (Figure~\ref{fig:kcurve}). The minimum's
location is stable ($K{=}20$ beats $K{=}40$ in six seeds of six, and the
$K{=}20$ mean lies below every seed's $K{=}1$ value), and its scale is the
one Section~\ref{sec:decomposition} predicts. A block's excess cumulative
variance grows with $K$ only until $K$ exceeds the correlation range several
times over; a three-to-five-ball range saturates near $K \approx 20$. The two
measurements were made by unrelated methods, and they agree. The closure is
also two-sided, as over-dispersion requires: block modes raise the simulated
WP in the hard bands and lower it in the easy ones.

Twenty-six percent is a floor, not an estimate of the whole effect. The block
bootstrap breaks dependence at every block boundary and captures none beyond
$K$; splicing longer segments does not help, because a long donor segment's
internal state diverges from the simulation's own state accounting, which is
why the curve goes flat rather than continuing downward past $K=20$.
Measuring the full contribution requires a generative dependent process (a
Markov-switching outcome model, say), which is no longer a diagnostic but a
different model, with a cost that Section~\ref{sec:tradeoff} takes up.

\paragraph{A control that failed.}
We also built what was intended as a heterogeneity-only arm: draw
cell-matched balls from \emph{random} positions of a single donor innings,
carrying the donor's latent conditions without its local ordering. We
report it because the failure mode is instructive. The arm produced a mean
absolute gap of 0.019, a better calibration than replaying whole real
trajectories achieves and one that no injection of innings-level
heterogeneity could produce; it also drifted from the shared marginals by
total variation 0.018, five times any block mode. The design is degenerate:
re-matching every ball to the simulation's evolving cell within one donor
turns the arm into a nearest-neighbour replay of empirical outcomes, an
estimator of $E[y \mid s]$ in disguise rather than an injection of
dependence. We flag it as a caution for constructive calibration experiments
of this kind; the mechanism split rests on the permutation decomposition of
Section~\ref{sec:decomposition}, which has no analogous failure mode.

\subsection{Where this leaves the model}\label{sec:leaves}

Assembled, Sections~\ref{sec:gap} and~\ref{sec:dependence} say the following.
The exact Markov WP is miscalibrated with a specific, out-of-sample-stable
shape: too extreme in both directions. Its one-ball marginals are essentially
perfect, so the error is the independence assumption itself. The dependence
that independence discards is identified: short-range sequential scoring
persistence of about three to five balls, four-fifths of it beyond anything
innings conditions explain, with wicket clustering ruled out. Injecting
that dependence, marginals fixed, closes a quarter of the calibration gap
at the range the decomposition measured, replicated across seeds.

The independence assumption is not incidental to the model. It is what makes
the state graph acyclic, the backward induction exact, and the win
probability a martingale: the property Section~\ref{sec:leverage}'s leverage
and WPA constructions stand on. The next section states the resulting
tradeoff precisely.

\section{The tradeoff}\label{sec:tradeoff}

The results of Sections~\ref{sec:gap}--\ref{sec:dependence} are one half of
a tradeoff, not a defect report; the other half is the machinery of
Section~\ref{sec:leverage}, and the two halves are the same assumption viewed
from opposite sides.

\subsection{The statement}\label{sec:statement}

Consider the three properties one might want from a single win-probability
object over the chase state $s = (b, w, r)$:
\begin{enumerate}
  \item \textbf{Exactness.} $V$ satisfies $V(s) = \sum_o p_o(s)\, V(s'_o)$
    identically, so that $V$ is a martingale by construction and leverage
    (the conditional MAD of $\Delta V$) and WPA (with
    $E[\mathrm{WPA} \mid s] = 0$) are exact bookkeeping rather than
    estimates.
  \item \textbf{Calibration.} $V(s) = E[y \mid s]$: among situations the
    model calls $p$, the chasing side wins fraction $p$.
  \item \textbf{The state is $(b,w,r)$.} No latent variables, no player
    identities, no history beyond the scoreboard.
\end{enumerate}

Sections~\ref{sec:gap}--\ref{sec:dependence} show empirically that these
three properties are jointly unsatisfiable for T20 chases. The construction
that delivers (1) assumes conditional independence of balls given $s$; the
true process carries short-range sequential scoring persistence beyond $s$
(Section~\ref{sec:decomposition}); therefore the constructed $V$ misses
$E[y \mid s]$ with the over-dispersed shape of Section~\ref{sec:signflip},
and injecting the missing dependence (the only remaining degree of freedom,
with marginals verified correct) moves $V$ toward calibration
(Section~\ref{sec:constructive}). Under (3), exactness and calibration
exclude each other. This is an empirical demonstration for this model class
and this sport, not a theorem; but the elimination in
Section~\ref{sec:elimination} gives it more force than a fitted-model
comparison would, because it identifies \emph{which assumption} fails and
confirms the mechanism constructively.

The tradeoff also does not dissolve from the other direction. One could fit $V^*(s) \approx E[y \mid s]$
directly (Section~\ref{sec:validation}'s gradient-boosted baseline is
this model) and obtain calibration by construction. But $V^*$ is a
projection of a dependent process onto a non-sufficient state, and along real
ball sequences it drifts: knowing the current state \emph{and the outcome
that produced it} shifts the expectation of $V^*$ at the next ball, because
the last outcome predicts near-future scoring
(Section~\ref{sec:decomposition}) in a way the state does not capture. After
a boundary, the true win probability sits above $V^*(s')$; after a dot-ball
stretch, below. Consequently $E[\Delta V^* \mid \text{path}] \neq 0$, WPA
computed on $V^*$ is not zero-mean given the situation (a batter would earn
``clutch'' credit partly for the surface's blind spot), and leverage loses
its interpretation as the dispersion of a fair game. A calibrated surface
over this state cannot host exact attribution either. The two desiderata do
more than trade against each other within one model: they end up in
different objects.

\subsection{The escape hatches, addressed}\label{sec:escapes}

\paragraph{Enrich the state.}
Property (3) is the scope of the claim, so the natural escape is a richer
state carrying the latent that drives the dependence. Two results constrain
this route. The obvious enrichment, how long the striker has been at the
crease (the direct observable proxy for ``set''), adds nothing: $+0.0004$
nats in the most flexible model we could fit (Section~\ref{sec:baselines}),
a null with a match-clustered CI tight around zero. And the decomposition
says why finding a better enrichment is hard: the residual dependence is
ball-adjacent with a three-to-five-ball range, surviving innings and
partnership demeaning, so it does not correspond to any slowly-varying
observable a scoreboard state could carry. It behaves like a hidden regime,
not a hidden covariate. We cannot rule out that some pre-ball-observable
enrichment restores both properties (the claim is scoped, not absolute), but
the one route the folklore suggests is measurably closed.

\paragraph{Model the dependence generatively.}
A Markov-switching outcome process (a hidden hot/cold scoring regime with
three-to-five-ball persistence) would recover calibration, and
Section~\ref{sec:constructive}'s 26\% is a floor on what it would gain. The
cost is structural. The state graph over $(b,w,r,\mathrm{regime})$ still
admits backward induction in principle, but the regime is unobservable: win
probability becomes an expectation over a filtered posterior, leverage
becomes model-dependent, and WPA attribution requires inferring each ball's
regime before crediting the player. Attribution stops being bookkeeping and
becomes inference, with all of its instability. This is the same fork taken
by the simulator line of \citet{davis2015}, who condition outcomes on player
identities: richer, better calibrated in expectation, and no longer exact.
The choice between the branches is real, and nothing in the data makes it
for you.

\paragraph{Recalibrate post hoc.}
A monotone map from $V$ to observed frequencies fixes the reliability diagram
and destroys the recursion: the mapped surface no longer satisfies
$V = \sum_o p_o V'$, so the martingale property is lost, and with it every
Section~\ref{sec:leverage} guarantee. This was the concrete choice
faced in Section~\ref{sec:model}: the recalibrated surface was measurably
better calibrated and was rejected for the leverage layer for this
reason. In light of Sections~\ref{sec:gap}--\ref{sec:dependence} that
rejection was not conservatism; it was forced.

\subsection{Practical resolution}\label{sec:resolution}

For applied work the resolution we adopt, and recommend, is two surfaces with
declared roles. \emph{Report} win probability from a calibrated surface
(here, even a one-feature logistic transfers across eras better than the
exact model). \emph{Attribute} (leverage, WPA, clutch splits) on the exact
martingale surface, and publish its known error bound alongside: residual
lag-1 autocorrelation $+0.037$, calibration gap up to 0.11 in the contested
band, both stable out of sample. The attribution rankings are robust to this
bias because it is shared across players and largely removed by the
state-conditional de-drifting of Section~\ref{sec:wpa}; the probabilities are
not, which is why they should come from the other surface. What one must not
do is take WPA totals computed on the exact surface and read them as
calibrated probability claims: the $+62.3$-win aggregate drift of
Section~\ref{sec:wpa} is that error, measured.

\section{Limitations}\label{sec:limitations}

\paragraph{Scope of data.}
One league (IPL, 2008--2026) and one innings type. Chases were chosen
deliberately, because a fixed target makes $(b,w,r)$ a complete scoreboard
description and the terminal conditions exact, but this means the tradeoff
is demonstrated here, not shown universal. First innings, other T20 leagues, and
the 50-over format are open. We expect the mechanism (scoring persistence
beyond a scoreboard state) to transfer; its magnitude may not.

\paragraph{The 26\% is a floor with an unmeasured ceiling.}
The block bootstrap breaks dependence at every block boundary and captures
nothing beyond its window; its curve saturating at $K \approx 20$ reflects
the method's reach as much as the effect's size. The full contribution of
sequential dependence to the calibration gap awaits a generative dependent
model, which is future work; per Section~\ref{sec:tradeoff}, that model is
a different object rather than a repair.

\paragraph{No clean simulator-side heterogeneity control.}
Our designed control degenerated into a nearest-neighbour replay
(Section~\ref{sec:constructive}), so the serial-vs-heterogeneity split rests
on the permutation decomposition alone. That test is exact and validated on
synthetics, but a second, independent constructive probe would be better, and
designing one that neither breaks marginals nor degenerates appears genuinely
hard.

\paragraph{Attribution simplifications.}
Run-outs are folded into the wicket outcome and credited to the bowler;
extras are folded into the batter's ball count (a wide faced increments
deliveries faced without consuming a legal ball). Both are disclosed
data-contract decisions rather than modelling claims, and neither touches the
calibration analysis, but WPA at the individual level inherits them.

\paragraph{Player-level uncertainty.}
Leaderboards carry match-clustered bootstrap intervals (2{,}000 replicates
per player, whole innings resampled; per-ball resampling would assume away
the dependence Section~\ref{sec:dependence} measured). They sharpen
Section~\ref{sec:leverage} in both directions. No adjacent pair in any
top-ten table is statistically separable (0 of 36 adjacent-rank difference
CIs exclude zero), so the rankings resolve roles rather than neighbouring
names, and the role contrast is decisive (mean LI $+0.69$, CI
$[+0.33, +1.08]$, top finisher against the lowest-leverage qualified
anchor). The headline clutch splits survive: Dhoni's is $+0.018$ per ball
(CI $[+0.004, +0.032]$), the Yadav inversion $-0.013$ (CI
$[-0.031, -0.001]$); five of seventeen qualifying batters have splits
excluding zero. The intervals condition on the fitted surfaces; the
model-level dependence bound of Section~\ref{sec:leverage-defs} sits on top
of them, and per-ball differences below about $\pm 0.005$ remain inside it.

\paragraph{In-play markets as a third surface.}
In-play betting markets price the same chase states on a strictly finer
information set: player identities, pitch behaviour, the live broadcast.
The dependence measured in Section~\ref{sec:dependence} makes a specific
prediction about any scoreboard-state price (after a boundary it sits
below the truth, after a run of dot balls above), and it is an open
empirical question whether market prices carry that persistence at the
right magnitude, over-shoot it, or ignore it. Testing whether
market-implied win probability behaves as a martingale with respect to
$(b, w, r)$ is a direct application of the machinery built here; it awaits
ball-level alignment of exchange odds with the ball-by-ball record, and we
leave it open.

\section{Conclusion}\label{sec:conclusion}

We built the exactly solvable win-probability model for T20 chases and used
its martingale structure to make leverage and clutch attribution exact,
confirming the folk hypothesis about finishers and death bowlers along the
way. We then turned the same rigour on the model itself and found its
probabilities systematically over-dispersed; localized the error; eliminated
the marginals; decomposed the residual dependence into a short-range
sequential scoring persistence that innings conditions mostly do not
explain, with wickets anti-clustering against expectation; and closed a
quarter of the calibration gap by injecting the dependence we had measured,
at the range we had measured it.

The finding is that these two threads are one. The independence assumption
that makes the win probability exact (acyclic state graph, one-sweep
backward induction, martingale attribution) is the same assumption whose
failure miscalibrates it. Within a scoreboard state, you may have exact
attribution or calibrated probability, and you must choose; we make the
choice explicit and give each surface its role. We suspect this tension is
not specific to cricket: any
sport whose win-probability models assume event-level independence over a
compact state, and whose real event streams carry momentum at any timescale,
faces the same fork, usually unstated. This paper states it for one sport,
measures both sides of it, and prices the ways out.

\bibliographystyle{plainnat}
\bibliography{refs}

\end{document}